\documentclass[10pt,a4paper,twoside]{article}
\usepackage{tabularx}
\usepackage{vmargin,fancyhdr, multicol,multirow,ifthen,
            graphicx,wrapfig,calc,dcolumn,apalike,setspace,
            boxedminipage, rotating,textcomp,picinpar,longtable,
            url,amsmath,amssymb,ogonek} 

\usepackage{upgreek}

\usepackage[tight]{subfigure}
\usepackage{imc2024}
\usepackage[T1]{fontenc}

\usepackage{color}

\usepackage{natbib}

\begin{document}
\SetPaperBodyFont

\begin{IMCpaper}{
\title{FRIPONMeter :  high resolution radiometer for fireballs}
\author{
M. Lebreton$^1$, 
H. Constans$^1$, 
F. Colas$^2$, 
P. Vernazza$^5$,
A. Malgoyre$^5$,
S. Bouley$^{8,4}$,
B. Zanda$^{9,4}$,
S. E. Anderson$^{6,4}$,
P. Shober$^{7,4}$,
J. L. Rault$^{2,4}$, 
D. Darson$^1$,
S. Anghel$^{2,3}$,
D. Baratoux$^{10}$,
M. Birlan$^{2,3}$,
S. Bouquillon$^2$,
J. Gattacceca$^{11}$,
S. Jeanne$^2$,
L. Jorda$^5$,
A. Lagain$^{11}$,
L. Maquet$^2$,
S. Sylla Mbaye$^{13}$,
J. Vaubaillon$^2$ and
P.M. Zanetta$^{12}$
 
\thanks
{$^1\,$LPENS, ENS, Univ. PSL, CNRS, Sorbonne Univ., France
$^2\,$LTE, Observatoire de Paris, Université PSL, Sorbonne Université, LNE, CNRS, Paris, France  
$^3\,$Astronomical Institute of the Romanian Academy, Bucharest, Romania
$^4\,$FRIPON and Vigie-Ciel Team, France
$^5\,$LAM - Pythéas - Marseille - France
$^6\,$OCA, CNRS, Laboratoire Lagrange, Nice, France
$^7\,$NASA Astromaterials Research and Exploration Science Division, Johnson Space Center, Houston, USA
$^8\,$Université Paris-Saclay, UMR CNRS 8148, GEOPS, Orsay, France
$^9\,$Institut de Minéralogie, MNHN, CNRS, Paris, 75005, France
$^{10}\,$GET, Université de Toulouse, CNRS  IRD, Toulouse, France
$^{11}\,$CEREGE, CNRS, AMU, IRD, PYTHEAS, Aix-en-Provence, France
$^{12}\,$LGL-TPE UMR5276 (CNRS, UCBL1, ENS Lyon, UJM), Saint-Étienne, France 
$^{13}\,$ITNA, Université Cheikh Anta Diop, Dakar, Senegal
 }}
                  
\abstract
{The FRIPON network has been operating for ten years, enabling the recovery of two meteorites in France and contributing to the recovery of five others across Europe. The fall of 2023 CX1 demonstrated that precise knowledge of the altitude of catastrophic fragmentation is essential for computing the dark flight and, consequently, the strewn field. This altitude can be determined through accurate measurements of surges in the light curve. Light curves also provide insights into the mechanical strength of meteoroid material and can help determine whether it has a brittle composition similar to that of cometary material. Although this is possible using camera data, the acquisition rate (30 fps) is often too low, and saturation issues arise for very bright events such as meteorite-producing fireballs. Radiometers offer an alternative method to overcome these limitations. The European Fireball Network was the first to install radiometers alongside fisheye cameras. This article presents a prototype designed for the FRIPON network that can also operate as a standalone instrument. Its main features include precise timing, a high acquisition rate (2 kHz), and an extended dynamic range, enabling observations from small fireballs detectable by FRIPON to extremely bright events capable of producing meteorites.}
}

\vspace*{-3\baselineskip}

\section{Fireball light curves}

Light curves of atmospheric entries provide access to several physical parameters of the incoming object. One of the most important is the initial mass, which can be estimated from the measurement of the total radiated light energy of the phenomenon. This quantity can be converted into kinetic energy using the luminous efficiency coefficient. The main difficulty with this method, which has been addressed by many authors \citep{2024Icar..40815807J,2024Icar..40715817L,2021A&A...652A..84D}, is the estimation of the luminous efficiency coefficient. This parameter depends on several factors, such as the initial velocity, entry angle, and composition of the meteoroid. According to these studies, the luminous efficiency can vary by up to two orders of magnitude. A dedicated study based on FRIPON data \citep{2021A&A...650A.159D} used masses derived from atmospheric deceleration measurements at high altitude to estimate the luminous efficiency\citep{2020A&A...644A..53C}.

Light curves can also be used to identify fragmentation events, which manifest as sudden bursts of brightness. These features provide information about the mechanical strength of the material, as the dynamic pressure responsible for fragmentation can be estimated from the altitude and the speed of these events. The precise timing of brightness bursts allows the fragmentation altitude to be determined using the kinematic model derived from the triangulation of camera observations \citep{2020A&A...644A..53C}. The study of the 2023 CX1 fall \citep{2025NatAs.tmp..190E} demonstrated that this measurement is essential for computing the dark flight and, consequently, determining the strewn field. For this analysis, light curves must therefore be accurately time-stamped and have sufficient temporal resolution.

\subsection{Previous developments}

Attempts to measure meteor light curves using long exposures date back to the first networks employing photographic plates in the Czech Republic which used rotating shutters for velocity measurement, and extend to more recent systems such as the Desert Fireball Network (DFN) in Australia, which uses digital cameras equipped with liquid-crystal shutters \citep{2012AuJES..59..177B, devillepoix2019}, similar to those used by the European Fireball Network \citep{2010pim9.conf...14B}.
In 2010, the All-Sky Video Meteor Network in Southern Ontario, Canada --- the ASGARD system \citep{2010JIMO...38...25B} --- was established. Using analog video cameras, it enabled the recovery of the Grimsby meteorite \citep{2011M&PS...46..339B}. Later, networks such as FRIPON \citep{2020A&A...644A..53C} adopted digital cameras. However, the light curves obtained by all these networks made for astrometry share common limitations, particularly in terms of acquisition speed and image saturation for bright events.

Radiometers provide an alternative approach to overcoming these limitations.A radiometer is a broadband photometric detector that measures the integrated intensity of incident light over a wide field of view and across a broad spectral range. For meteor observations, it requires a high temporal resolution.
For this reason, the European Fireball Network installed radiometers alongside fisheye cameras \citep{2002ESASP.500..257S}. These radiometers were initially based on photomultipliers operating at 500~Hz and later at 5~kHz. Subsequently, radiometers based on photodiodes were developed \citep{2015pimo.conf..180V,2019pimo.conf..123B,2020pimo.conf..112R,app15052828}, but none have been deployed on a scale comparable to that of the European Fireball Network.

\begin{figure}[t!]
\includegraphics[width =0.9\columnwidth]{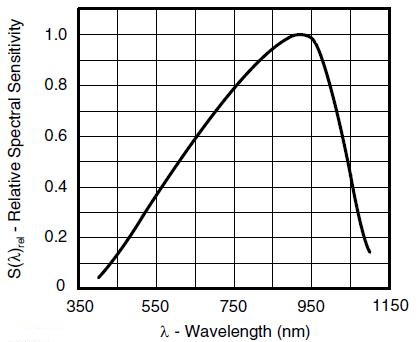}%
\vspace*{3pt}%
\caption{Photodiode spectral sensitivity.}
\label{Figure21}
\end{figure}

\subsection{Radiometer versus camera}

It is important to combine these two types of receivers because cameras allow photometric calibration using background stars, which radiometers do not. However, radiometers have a much higher temporal resolution and do not saturate during very bright events. By combining the two approaches, it is possible to obtain a light curve that can be used to measure the total luminous flux and, subsequently, the initial kinetic energy via the luminous efficiency coefficient. Finally, from the kinematic model obtained by triangulating camera observations from several stations, a photometry-based estimate of the initial mass can be derived. For non-saturated large bolides, this provides one of the best estimations of the pre-entry mass \citep{2021MNRAS.508.5716A}. 
The DFN tested the use of high-speed digital cameras (470~fps) and showed that they can be used as stand-alone instruments for both trajectory determination and unsaturated light-curve measurements \citep{2026arXiv260219447G}. Radiometer measurements, however, cover the entire sky and can have an even higher sampling rate and sensitivity. This also means that they are therefore more affected by variations in sky brightness (diffuse sky background, Moon, aircraft, light pollution, etc.). Although it can produce false detections, they are generally easier to filter out. \citet{2023MNRAS.518.2810A} showed that machine-learning techniques can filter more than 98\% of false positives in FRIPON-MOROI data, surpassing human performance.
 
\subsection{Specification aims}\label{aims}

Since the photometer is intended for fireball detection networks, and more specifically for FRIPON, the detection threshold must be close to magnitude~$-3$. For the brightest observable event, we based our calculations on the Chelyabinsk fireball \citep{2013Natur.503..238B}, which had an absolute magnitude of $-28$, close to that of the Sun ($-27$). However, for such bright events, it is not necessary to obtain unsaturated data from stations located close to the fall; more distant stations can easily be used. Therefore, this criterion on the brightest event observable without saturation is not mandatory. 
Based on previous experiments, in particular those conducted by the Czech team \citep{2002ESASP.500..257S}, an acquisition rate of a few kHz appears sufficient to study fragmentation.

\section{Design presentation}
 
\subsection{Photodiode measurement principle}
As in \citep{2020pimo.conf..112R}, we chose to use the same low-cost photodiodes to achieve a wide dynamic range of measurements.The Vishay Silicon PIN photodiode is specified with a 5~$\upmu$A/W/m$^2$ sensitivity at 950~nm, and its estimated sensitivity to solar spectral irradiance is 3.5~$\upmu$A/W/m$^2$. The spectral bandwidth sensitivity is that of a PIN silicon diode, with a maximum in the near infrared. The half-sensitivity range spans from 600~nm to 1050~nm (Figure~\ref{Figure21}).

The BPW34 photodiode is suitable for both low- and high-level measurements owing to different bias schemes (Figure~\ref{Figure1}):

\begin{itemize}
    \item Low-level measurements are performed by parallelizing 65 photodiodes operating at zero bias voltage (Operation in photovoltaic mode minimises the dark current).
    \item High-level measurements are performed by a single photodiode using up to 10~V bias voltage (Operation in photoconductive mode improves the linearity).
\end{itemize}

\begin{figure}[t!]
\centering
\includegraphics[width=\columnwidth]{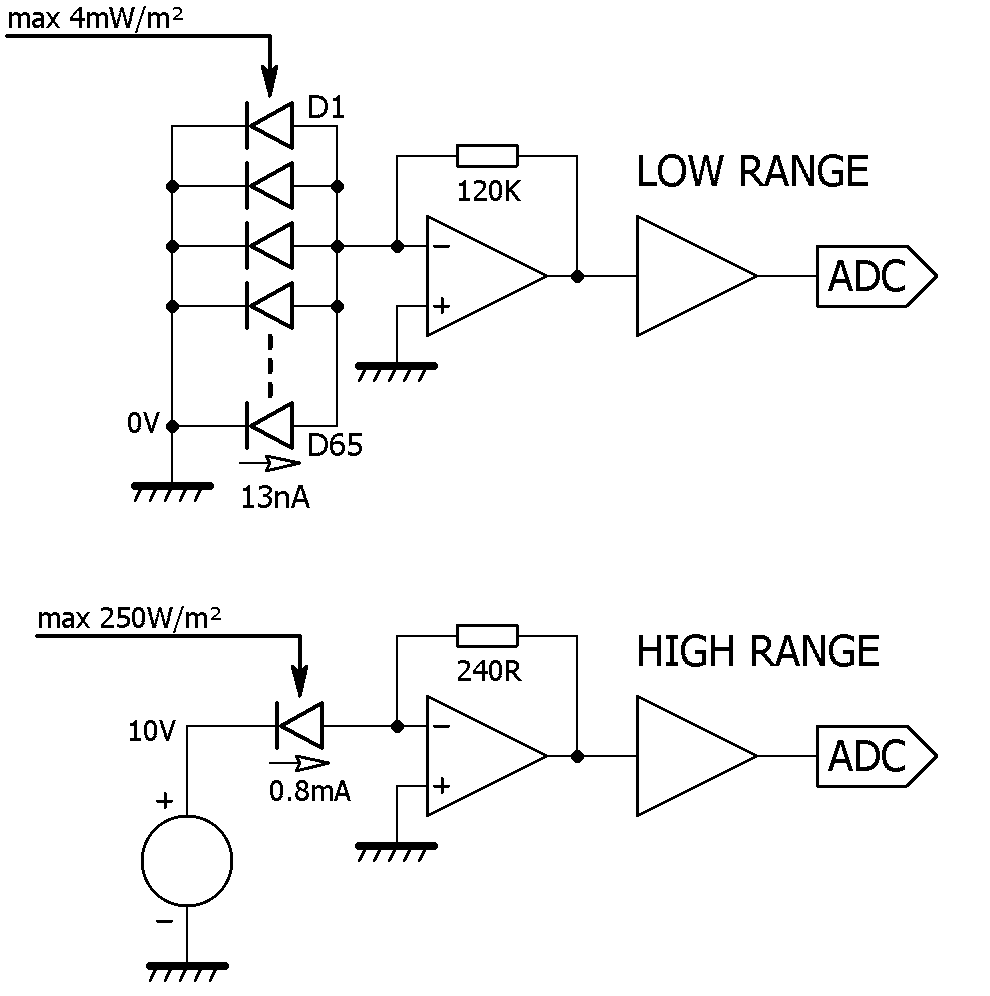}%
\vspace*{3pt}%
\caption{Photodiode amplifier schematic.}
\label{Figure1}
\end{figure}

In addition, the design of the amplifiers and ADC converters requires careful attention to optimize the signal-to-noise ratio. 
The sensitive area of a single photodiode is 7.5~mm$^2$, while the total area of the complete matrix of 67 photodiodes is 500~mm$^2$. The photodiodes are mounted on a 41$\times$41~mm horizontal flat surface, protected by a glass screen, and occluded at 45$^\circ$ by the aluminum housing window (Figure~\ref{Figure2}).

The sensitivity of the photodiodes decreases to 50\% at an elevation angle of 25$^\circ$. This value is adopted for the complete radiometer, based on the average vignetting caused by the metal housing window. The single diode covering the upper range, located at the centre of the matrix, is not subject to any geometrical limitation (Figure~\ref{Figure2}). Furthermore, the photodiode matrix is shielded from electrical noise by a grounded grid, which slightly attenuates incoming light. 

\begin{figure}[t!]
\centering
\includegraphics[width =\columnwidth]{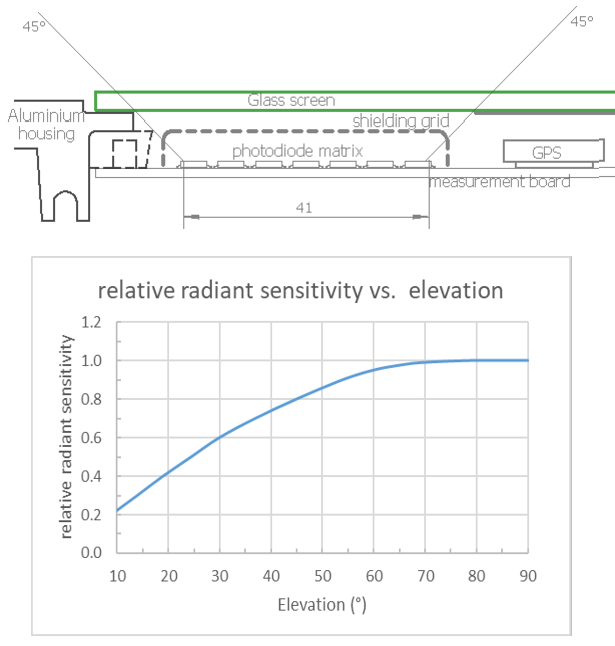}%
\vspace*{3pt}%
\caption{Photodiode matrix window profile.}
\label{Figure2}
\end{figure}

\begin{table*}[t!]
	\caption{Measurement system specifications (ambient temperature = 25$^\circ$C, unless otherwise specified).}%
	\label{table1}
	\vspace*{6pt}
	\centering
	\def\0{\ensuremath{\phantom{0}}}%
	\def\1{\ensuremath{\phantom{.0}}}%
{\footnotesize \begin{tabular}{lll}
		\hline
		\textbf{Parameter} & \textbf{Test condition} & \textbf{Value} \\
		\hline
		ADC converter resolution & either range & 16 bits \\
		Differential linearity & either range & $\pm$ 0.6 ADU (i.e., LSB)\\
		High range resolution & Sun light spectral irradiance at sea level & 3.87E-03 W/m\textsuperscript{2}/ADU  \\
		Medium range resolution & Sun light spectral irradiance at sea level & 1.56E-05 W/m\textsuperscript{2}/ADU  \\
		Low range resolution & Sun light spectral irradiance at sea level & 6.34E-08 W/m\textsuperscript{2}/ADU  \\
		Dark current + Bias current offset & Darkness, any range & 530 ADU \\
		Offset thermal drift & $-25^\circ$C / $+50^\circ$C & 2 ADU/$^\circ$C \\
		H. range / M. range gain ratio & nominal value & 249 \\
		M. range / L. range gain ratio & nominal value & 245 \\
		Gain thermal drift & $-25^\circ$C / $+50^\circ$C (BPW34 specification)& 0.1\%/$^\circ$C \\ 
		Peak-to-peak noise high range & darkness, 100 ms acquisition & 1 ADU \\
		Peak-to-peak noise medium range & darkness, 100 ms acquisition & 2 ADU \\
		Peak-to-peak noise low range & darkness, 100 ms acquisition & 4 ADU \\
		Internal µc filter cutoff frequency & $-3$ dB & 0.8 kHz \\
		Settling time at 99\% & Light step test & 1 ms \\
		Data recording rate & & 2 kHz \\
		\hline
	\end{tabular}}
\end{table*}

\subsection{Measurement range}

Maintaining both amplitude resolution and signal-to-noise ratio throughout the entire dynamic range is achieved by three simultaneous acquisition channels: 
high range = 250~W/m$^2$,
medium range = 1~W/m$^2$, 
and lower range = 4~mW/m$^2$.
The photometer continuously records 16-bit raw data from three simultaneous analog-to-digital converters. Post-processing of the data is necessary to select the appropriate range or to combine multiple ranges when the dynamic range exceeds that of a single channel. Since the gains and offsets of the three channels vary with temperature and aging, the processing will be performed independently for significant events, such as meteorite falls.

Post-processing of the data is also necessary to remove the continuous sky background in order to calculate the instrumental magnitude of each event. Finally, these data are adjusted to those provided by the cameras to obtain a light curve calibrated in absolute magnitude.
The operating range table (Figure~\ref{Figure20}) summarizes the key performance characteristics of this design.

\begin{figure}[b!]
	\centering
\includegraphics[width =\columnwidth]
{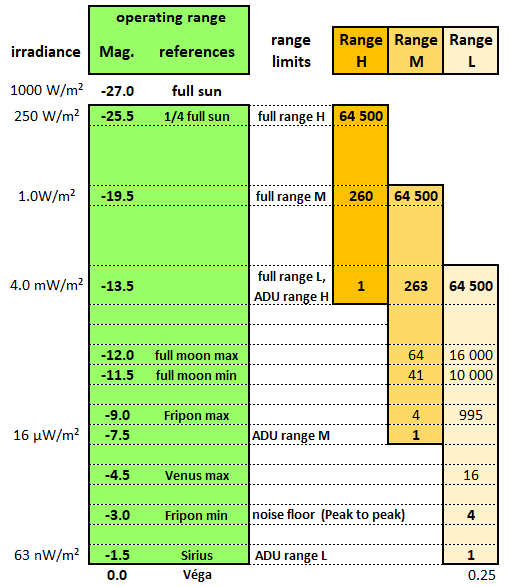}%
\vspace*{3pt}%
\caption{Operating range.}
\label{Figure20}
\end{figure}

The main performance features are as follows:
\begin{itemize}
    \item The low range is sufficiently sensitive and broad to monitor faint events, even when the full moon increases the sky background light.
    \item The high range, with a maximum level up to 25\% of full sunlight, is sufficient to fully capture any large event.
    \item The noise-free dynamic range reaches $1.0 \times 10^{9}$, corresponding to a 30-bit noise-free resolution.
    \item The specified noise amplitude is peak-to-peak over the full bandwidth of 0.8~kHz. Additional post-processing filtering allows the user to increase the effective resolution from 30~bits up to $4.0 \times 10^{9}$ (32~bits), based on the rms noise level.
    \item To avoid negative signals while accounting for the dark current, the input amplifiers are biased with an offset of 500 ADU.
\end{itemize}
 
The maximum observable magnitude is $-25.5$, which does not reach the target of $-27$. However, as explained in Section~\ref{aims}, this is sufficient for an extended network. This saturation limit can be compared with those of different networks: DFN at magnitude~$-15$ \citep{2021PASA...38...16N}, FRIPON at magnitude~$-9$ \citep{2020A&A...644A..53C}, SonataCO at magnitude~$-6$, and GMN at magnitude~0 \citep{2021MNRAS.506.5046V}. Apart from DFN and FRIPON, the latter networks are primarily dedicated to meteor observations rather than to fireballs.


\subsection{Hardware and connection design}

The radiometer comprises six internal functional blocks (Figure~\ref{Figure3}):

\begin{enumerate}
    \item 67-photodiode matrix;
    \item 3-range measurement circuit;
    \item temperature sensor;
    \item GPS/GNSS receiver for accurate time stamping;
    \item digital circuitry with microcontroller and Ethernet controller;
    \item isolated DC/DC converter and voltage regulators.
\end{enumerate}

The radiometer can be operated using one of two available external connections:

\begin{itemize}
    \item The optimal performance is achieved via the Ethernet PoE (RJ45) connection, which provides both network access and an isolated power supply. In this mode, the radiometer is integrated into the existing FRIPON camera system network (Fig.~\ref{Figure13}).
    \item Alternatively, a USB connection is available for simple standalone operation with a nearby PC.
\end{itemize}

\begin{figure}[t!]
\centering
\includegraphics[width = 0.75\columnwidth]{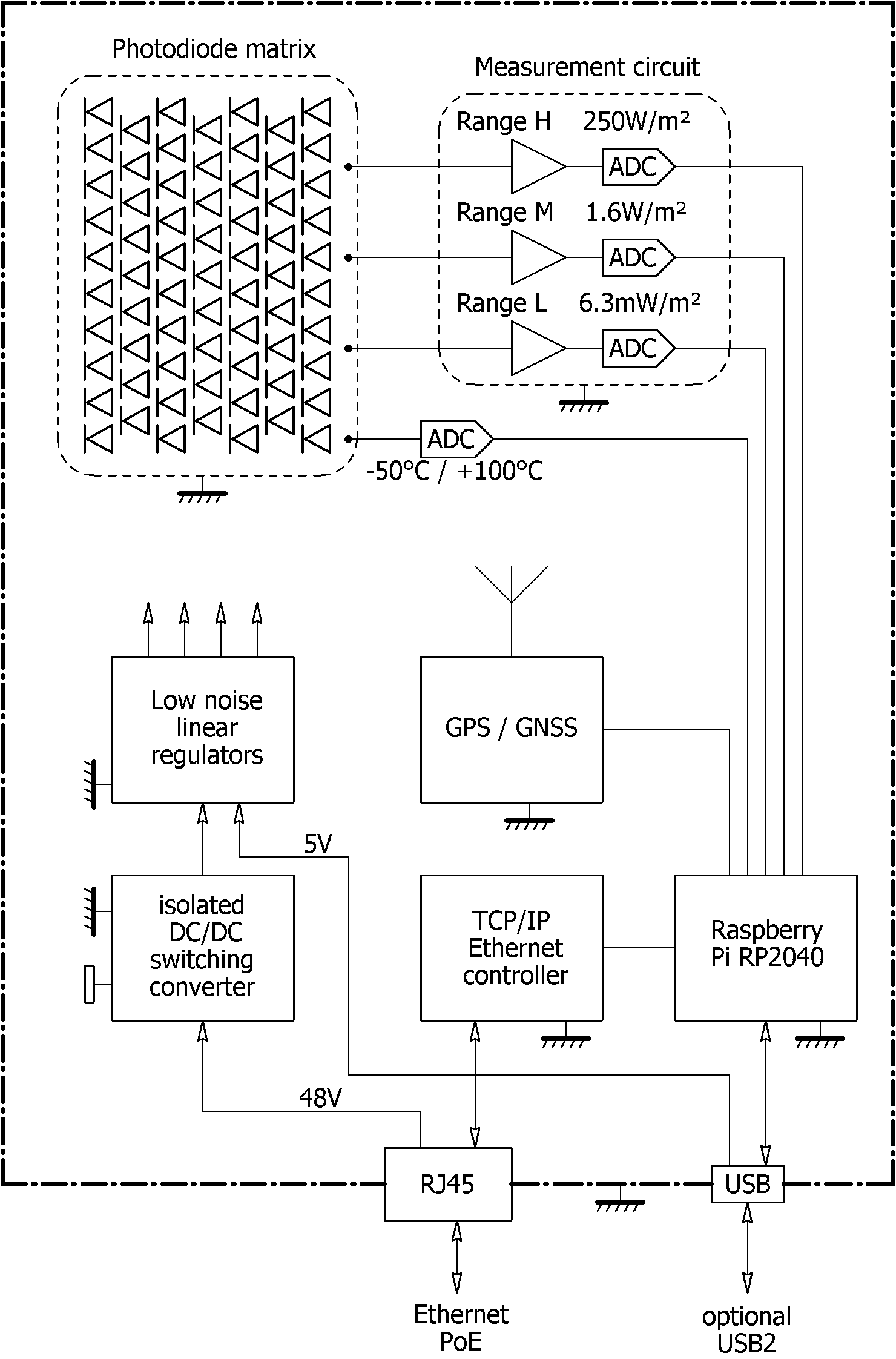}%
\vspace*{3pt}%
\caption{Hardware description.}
\label{Figure3}
\end{figure}

Whatever the external connection scheme, the low-level measurement requirements critically influence the hardware design. 
The lowest irradiance resolution, 63~nW/m$^2$ (magnitude~$-1.5$), corresponds to a photodiode current of only 0.2~pA. 
To enable accurate measurement of such a low-level current, several optimizations have been implemented:

\begin{itemize}
    \item The 65-photodiode matrix increases the amplifier input current to 13~pA. However, the capacitive input witch is coupled with the surrounding  electric field also increases noise proportionally with the total surface area of the matrix.
    \item The input stage employs a high-quality, low-noise amplifier and low-value gain resistors.
    \item Internal digital filtering suppresses high-frequency switching noise from the power supply.
    \item All circuitry is shielded using common ground planes.
    \item The photodiode matrix itself is shielded with a grounded metallic front grid.
    \item The radiometer shielding ground must be connected to the mains earth via an FTP-shielded RJ45 cable and a compatible PoE switch.
\end{itemize}
This final optimization is mandatory to ensure the radiometer's performance. Conventional PoE switches with 48~V~DC power supplies are designed for applications where electrical noise coupling is relevant only in terms of EMC standards and operator safety. Accordingly, the metal housing and shielding are not connected to the mains earth. Furthermore, the internal transformer isolation of low-cost 48~V~DC power supplies is poor: stray capacitances can drive significant 50~Hz AC currents to the connected devices (i.e., common-mode currents). If unmitigated, these currents can leak through the RJ45 cable, the photodiode matrix, and its capacitive surface to earth; during a darkness test, a 50~Hz input noise of 400~ADU was measured in the lower range. To prevent this, an additional cable, connected as close as possible to the PoE switch outlet, grounds the metal housing and shielding of the PoE switch, as illustrated in Figure~\ref{Figure13}. When using the optional non-isolated USB connection for standalone operation with a local PC, it is important to ensure that the PC is itself properly grounded to the mains earth.

\begin{figure}[t!]
\centering
\includegraphics[width = 0.7 \columnwidth]{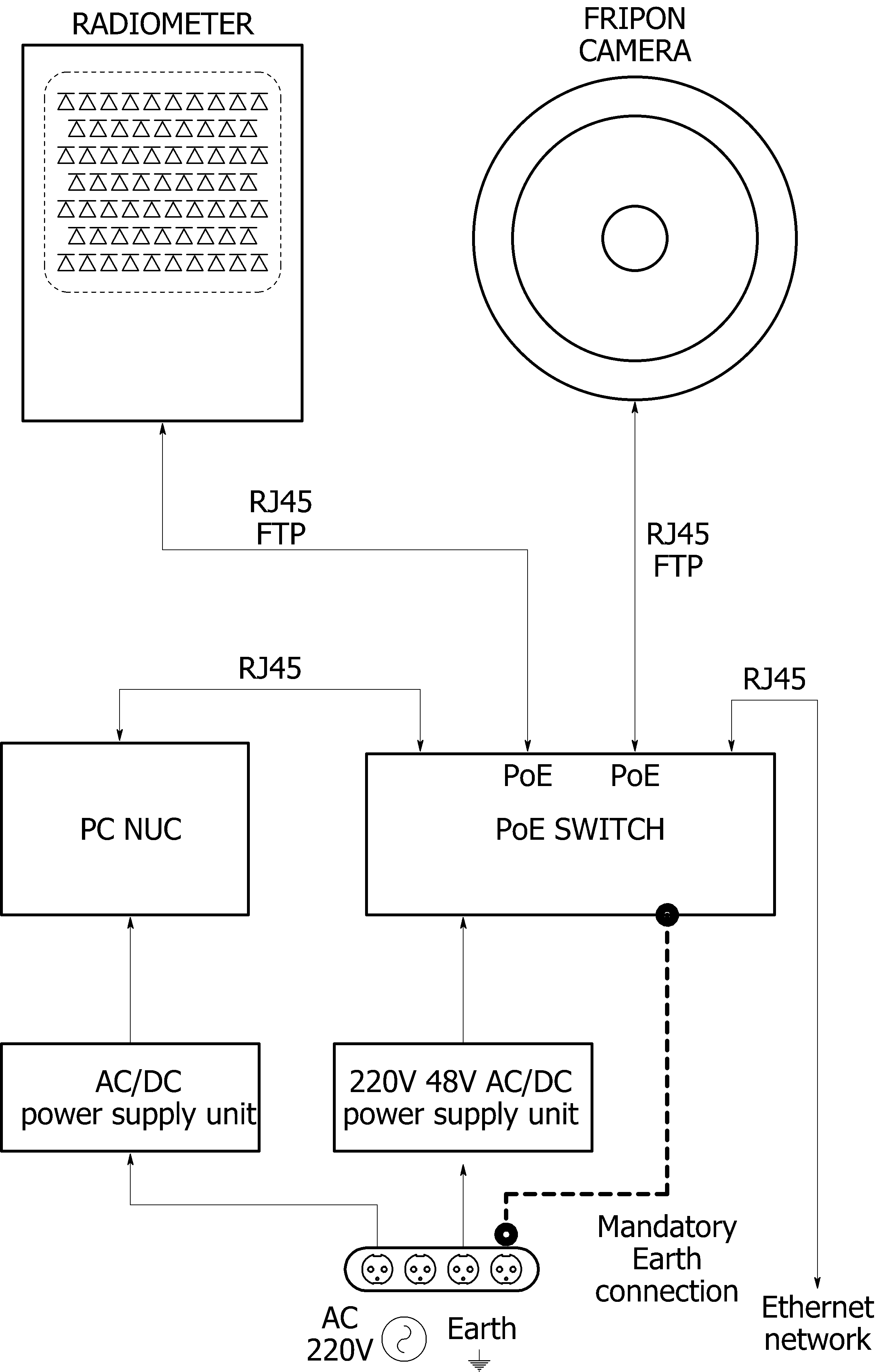}%
\vspace*{3pt}%
\caption{Ethernet PoE FRIPON network.}
\label{Figure13}
\end{figure}

\subsection{Mechanical design}

The radiometer prototype was designed and tested using a waterproof plastic enclosure with a 150 $\times$ 85 $\times$ 80~mm polycarbonate transparent lid, as illustrated in Figure~\ref{Figure15}. The next version will use a waterproof aluminum 159 $\times$ 86 $\times$ 60~mm enclosure fitted with a glass screen integrated into the lid, as shown in Figure~\ref{Figure16}.

\begin{figure}[htb]
\centering
\includegraphics[width = 0.7 \columnwidth]{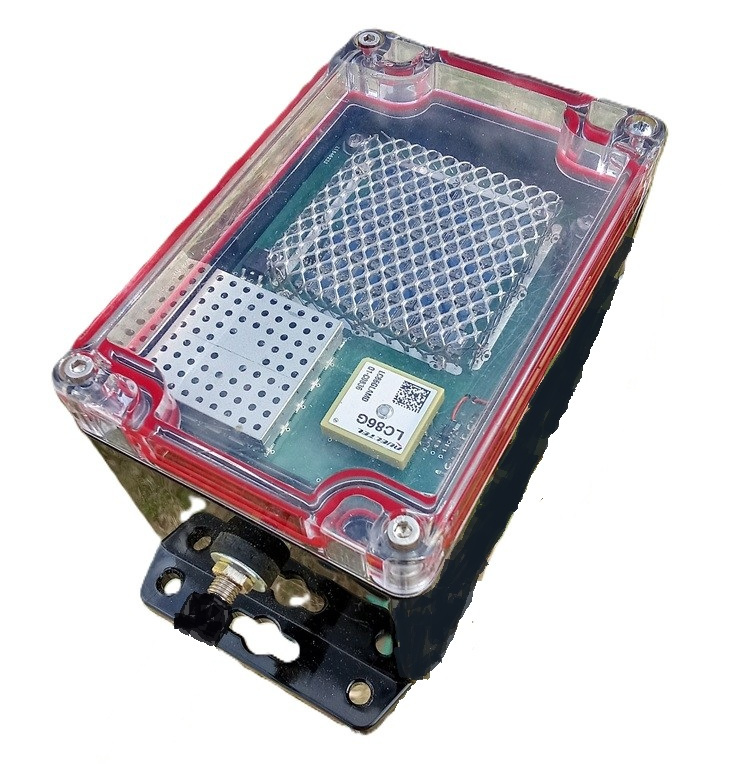}%
\vspace*{3pt}%
\caption{Radiometer prototype A.}
\label{Figure15}
\end{figure}

\begin{figure}[htb]
\centering
\includegraphics[width = \columnwidth]{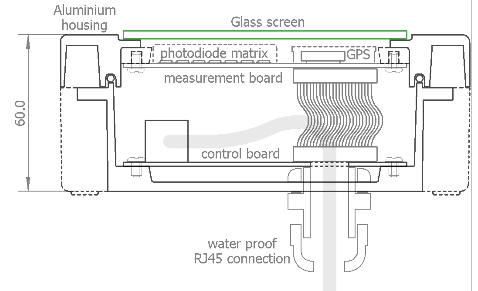}%
\vspace*{3pt}%
\caption{Radiometer version B.}
\label{Figure16}
\end{figure}

\subsection{Software and data storage}

The photometer onboard computer is based on a Raspberry Pi Pico RP2040 microcontroller. Its embedded firmware initializes a TCP server, performs ADC readings every 20~$\upmu$s, and applies a first-order infinite impulse response (IIR) low-pass filter to these samples. The radiometer digital specifications are shown in Table~\ref{Radiometer digital specifications}. One second of data is stored in a 2k-sample buffer, corresponding to a 500~$\upmu$s sampling interval. When a remote computer is connected, the acquisition buffer is transmitted every second. Measurement values are sent as raw hexadecimal numbers from the 16-bit unsigned ADC output. Each buffer is labeled with a GPS time stamp, the photometer temperature, and a sample count. Data are lost if no remote computer is connected. A TCP client on the same network can continuously read the data frames. Table~\ref{Data storage specifications} summarizes the storage specifications without size optimization.

\begin{table}[!ht]
\caption{Radiometer digital specifications}%
\label{Radiometer digital specifications}
\vspace*{6pt}
\centering
\def\0{\ensuremath{\phantom{0}}}%
\def\1{\ensuremath{\phantom{.0}}}%
{\footnotesize	\begin{tabular}{lccc}
\hline 
\textbf{Time} & \textbf{Memory size}\\
\hline
Internal sampling frequency & 50 kHz for each range\\
Internal digital filter cutoff frequency & 800 Hz\\
Quantization & 2 kHz\\
Recording time stamping resolution & $\pm$ 500~$\upmu$s \\
Data frame size jitter & $\pm$ 1 sample\\
Max. internal clock frequency drift  & 50 ppm/$^{\circ}$C \\
\hline
\end{tabular}}
\end{table}

\begin{table}[!ht]
\caption{Data storage specifications.}%
\label{Data storage specifications}
\vspace*{6pt}
\centering
\def\0{\ensuremath{\phantom{0}}}%
\def\1{\ensuremath{\phantom{.0}}}%
{\footnotesize	\begin{tabular}{lccc}
\hline 
\textbf{Time} & \textbf{Data} & \textbf{Content} & \textbf{Memory size}\\
\hline
1 s & frame & 2k samples $\times$  3 ranges & 24 kB\\
1 min & file & 60 frames & 1.4 MB\\
1 day & directory & 1440 files & 2.1 GB\\
1.3 year & disk drive & 480 days & 1 TB\\
\hline
\end{tabular}}
\end{table}


\section{Tests on the sky}

\subsection{2025 Perseid meteor shower}

We took advantage of the 2025 Perseid meteor shower to organize an observation campaign at the Pic du Midi observatory, even though this year was not favorable due to a very bright moon. With detection statistics being low for the FRIPON network (two or three detections per week per station), it was important to take advantage of one of the largest meteor showers of the year. We chose the Pic du Midi observatory (Figure~\ref{Figure33}) because it is equipped with a FRIPON camera which allows for the photometric calibration of FRIPOMeter .

\begin{figure}[htb]
\includegraphics[width = \columnwidth]
{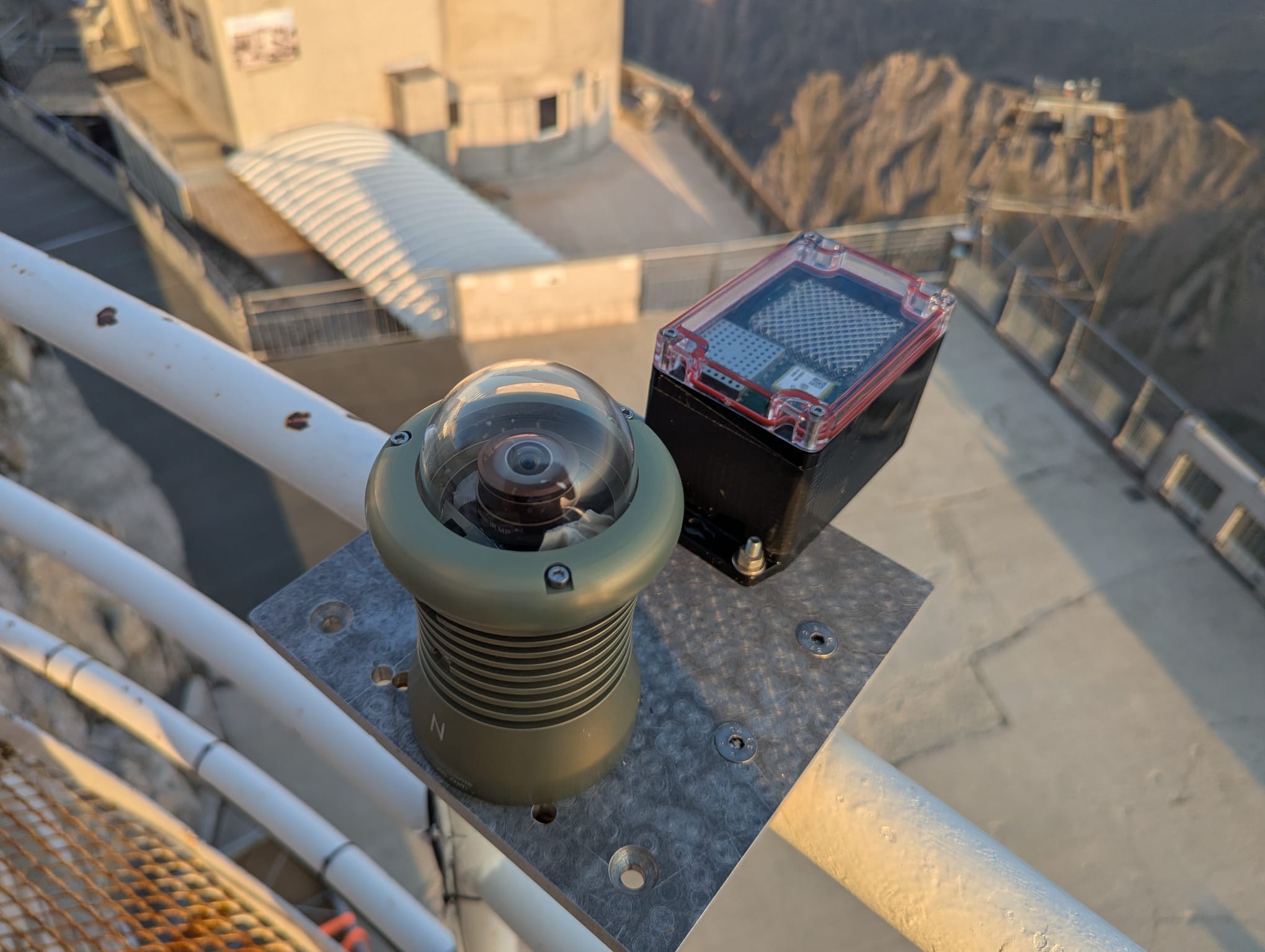}
\vspace*{3pt}
\caption{FRIPON camera and radiometer at Pic du Midi observatory.}
\label{Figure33}
\end{figure}

\subsection{One event data processing example}

Data frames from the radiometer are continuously recorded by the remote PC, which also controls the FRIPON camera. Each frame contains the raw acquisition data from the three simultaneous ranges over a one-second interval.

\begin{itemize}
    \item When FRIPON detects a fireball near the Pic du Midi, the corresponding data frame is retrieved for further analysis.
    \item The electronic offset plus sky background are estimated by averaging the signal over a 10-second interval before and after the event. A light curve in arbitrary units (ADU) is then extracted and calibrated using the FRIPON pipeline to derive absolute magnitudes.
    \item A low-pass digital filter may be applied in post-processing to enhance the signal-to-noise ratio for faint events with slow temporal variations.
\end{itemize}
As an example, a fireball was recorded on 2025-08-12 at 01:53:29~UT from Pic du Midi (Figure~\ref{Detection1}).
The background sky brightness was 8275~ADU (low range), corresponding to nearly full-moon conditions (magnitude~$-11.3$).
This demonstrates that even under full-moon conditions, the radiometer is capable of recording fast and faint events  at its maximum temporal resolution.
The detection threshold is magnitude~$-2.5$, in good agreement with the instrument specifications.

\begin{figure}[htb]
\includegraphics[width = \columnwidth]
{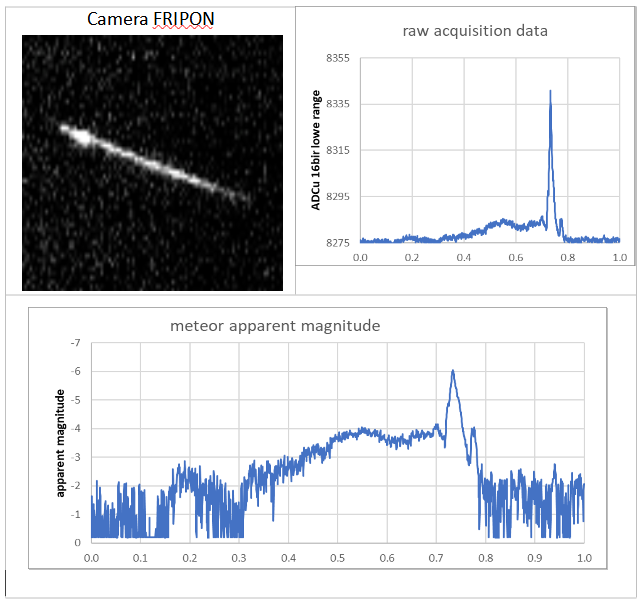}%
\vspace*{3pt}%
\caption{Fireball detected on 2025-08-12 at 01:53:29~UT observed from Pic du Midi. Top right: raw linear data, bottom: absolute magnitude data calibrated with FRIPON camera data.}
\label{Detection1}
\end{figure}

\subsection{Calibration on sky}

A single unsaturated bright event is used for the calibration: peak magnitude $-8$, elevation $64^\circ$, altitude $96$~km, and distance $105$~km.
The gain of the lower dynamic range is computed by comparing data from the FRIPON camera and the radiometer (Fig.~\ref{Figure8}), both located at the same site. 

The processing steps are as follows:

\begin{enumerate}
    \item Data centered around the event are retrieved.
    \item A digital post-processing filter with a cutoff frequency of 80~Hz and a settling time of 10~ms is applied to reduce the readout noise.
    \item A constant level of 2545~ADU, corresponding to both the electronic offset and the sky background (approximately magnitude $-10$), is subtracted.
    \item The radiometer gain is fitted to obtain the best alignment between the camera and radiometer light curves.
\end{enumerate}

The radiometer absolute magnitude of the event is computed using Equation~\ref{eq:gain-mag}, the gain is ajusted with FRIPON data:
\begin{equation}
    mag = -2.5 \times \log \left( \frac{ADU}{gain} \right) 
\label{eq:gain-mag}
\end{equation}

The difference between the factory gain value of 0.275 (computed using the estimated photodiode sensitivity to solar spectral irradiance of 3.5~$\upmu$A/W/m$^2$) and the gain calibrated on the sky, 0.25, is less than 9\%.

\begin{figure}[htb]
\includegraphics[width = \columnwidth]{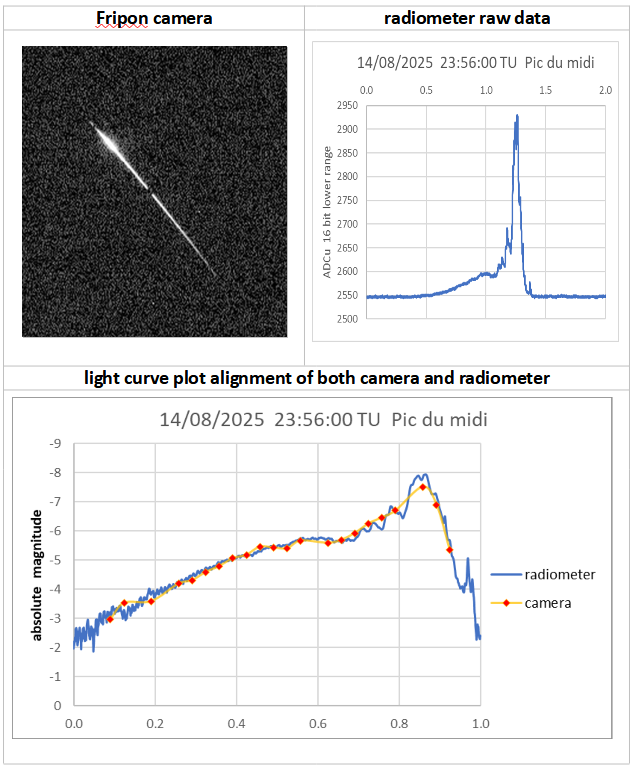}%
\vspace*{3pt}%
\caption{Photometric calibration.}
\label{Figure8}
\end{figure}
Since no comparison between the camera and the radiometer is available at high signal levels, the gains of the medium and high ranges are calibrated relative to the low range gain. Twilight sky brightness is used to obtain several measurement points across the transitions between low and medium ranges, and then between medium and high ranges. These transitions provide a sufficiently wide overlap to allow a linear interpolation over several points and thus determine an accurate gain ratio. The difference between the gain ratios (high/medium or medium/low) derived from the factory specifications (at 25$^\circ$C) and those measured on the sky is less than 1.5\%.

\subsection{Interference events}

The main source of light pollution is public lighting, which is modulated at 100~Hz (twice the grid frequency). Unfortunately, removing this interference is not possible, as it lies within the frequency range of scientific interest for meteor observations.This issue motivates the installation of both the camera and the radiometer at sites light pollution is low.
Several other types of events can bias the fireball monitoring; fortunately, they mainly produce short, high-speed spikes that are easy to distinguish from a meteor light curve. Moreover, these short events provide a convenient way to test the response time of the radiometer over a wide range of amplitudes. An example recording of a thunderstorm lightning flash during daylight, obtained using the high range (maximum $250~\mathrm{W\,m^{-2}}$), is shown in Fig.~\ref{Figure9}. Because the light pulse is shorter than the sampling period, the recorded pulse shape corresponds to the response of the internal 800~Hz low-pass filter (only three samples $\times$ 0.5~ms $= 1.5$~ms).

\begin{figure}[htb]
\includegraphics[width = \columnwidth]{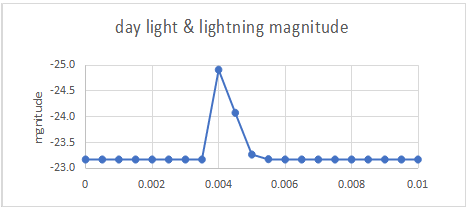}%
\vspace*{3pt}%
\caption{Apparent magnitude of a thunderstorm lightning recorded during daylight.}
\label{Figure9}
\end{figure}

An example of a nearby airliner with periodically flashing lights is shown in Fig.~\ref{Figure10}. The duration of each light pulse is shorter than the radiometer sampling period of 0.5~ms. The aircraft periodic pattern, with several colors and orientations, is recorded with varying amplitudes. The burst record lasts for several minutes, corresponding to the time required for the aircraft to cross the sky. A small meteor event occurs simultaneously, demonstrating that this type of interference can be easily removed from the final data.

\begin{figure}[t!]
\includegraphics[width = \columnwidth]{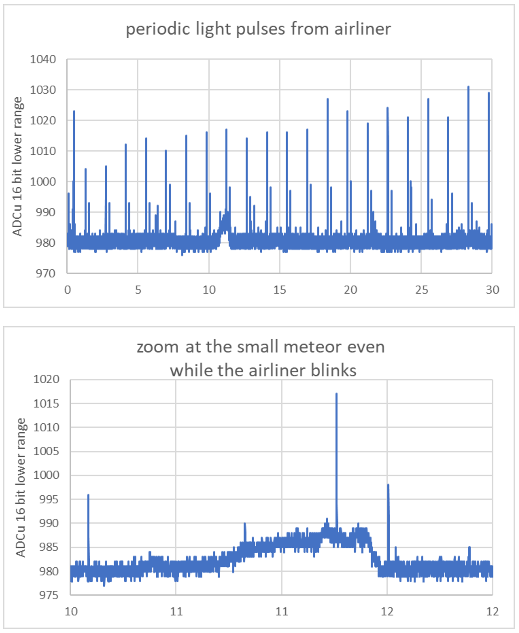}%
\vspace*{3pt}%
\caption{Periodic light pulses from a nearby airliner. ADU as a function of time.}
\label{Figure10}
\end{figure}

\section{Next steps and conclusions}
As the technical characteristics of this radiometer prototype comply with the specifications, its overall design (photodiodes, analog-to-digital converters, and PoE power supply) will be retained for the development and manufacture of a new prototype. However, several modifications will be introduced. In particular, a USB control interface will be added, allowing more flexible use in networks that do not employ the FRIPON configuration \citep{2020A&A...644A..53C}, which relies on a PoE connection. 
In addition, PoE control components will be integrated into the digital board, resulting in a simplified architecture with only two electronic boards: one dedicated to the photodiodes with analog-to-digital conversion and GPS, and a second handling power supply and communication with the control PC. Finally, a new metal enclosure with a glass window will be used in order to maximize the operational lifetime of the radiometer.

During the \textit{Pic du Midi} observational run, events occurring at distances greater than 200 km were detected. Because the FRIPON network is very dense (median inter-station distance $\sim$80\,km), radiometeres are not required at every station. One instrument every 160\,km is sufficient. We therefore plan to manufacture approximately 100 units to equip 25 stations in France, 50 worldwide, and keep 25 as spare units.
The manufacturing plans will be released as open-source hardware, along with the software, to allow the FRIPONMeter to be freely replicated. We plan to manufacture and test this second prototype in 2026 and begin mass production in 2027. The estimated cost will be approximately €300.

\bibliographystyle{imo2}
\bibliography{FripoMeter}

\end{IMCpaper}
\end{document}